\documentclass[aps,prl,reprint,superscriptaddress]{revtex4-1}

\usepackage{graphicx}
\usepackage{dcolumn}
\usepackage{bm}
\usepackage{color}
\usepackage[french]{babel}
\preprint{APS/123-QED}

\begin{document}

\title{Field induced anisotropic cooperativity in a magnetic colloidal glass}
\author{E. Wandersman}
\email{elie.wandersman@upmc.fr}
 \affiliation{
 Sorbonne Universit\'es, UPMC Univ Paris 06, UMR 8234, Laboratoire PHENIX -  CNRS - UPMC - ESPCI, Bo\^ite 51, 4 place Jussieu, F-75005, Paris, France}

 \affiliation{   
Sorbonne Universit\'es, UPMC Univ Paris 06, UMR 8237, Laboratoire Jean Perrin, CNRS - UPMC, Bo\^ite 114, 4 place Jussieu, F-75005, Paris, France}

\author{Y. Chushkin}
\email{chushkin@esrf.fr}
 \affiliation{
 European Synchrotron Radiation Facility - 6 rue J. Horowitz BP 220, 38043
Grenoble Cedex 9 France}

\author{ E. Dubois}
 \affiliation{
  Sorbonne Universit\'es, UPMC Univ Paris 06, UMR 8234, Laboratoire PHENIX -  CNRS - UPMC - ESPCI, Bo\^ite 51, 4 place Jussieu, F-75005, Paris, France}

\author{V. Dupuis}
 \affiliation{
 Sorbonne Universit\'es, UPMC Univ Paris 06, UMR 8234, Laboratoire PHENIX -  CNRS - UPMC - ESPCI, Bo\^ite 51, 4 place Jussieu, F-75005, Paris, France}

\author{A. Robert}
 \affiliation{
 SLAC National Accelerator Laboratory, Linac Coherent Light Source, 2575
Sand Hill Rd, Menlo Park CA 94025 - USA}

\author{R. Perzynski}
 \affiliation{
 Sorbonne Universit\'es, UPMC Univ Paris 06, UMR 8234, Laboratoire PHENIX -  CNRS - UPMC - ESPCI, Bo\^ite 51, 4 place Jussieu, F-75005, Paris, France}

\date{\today}
\begin{abstract}
The translational dynamics in a repulsive colloidal glass-former is probed by time-resolved X-ray Photon Correlation Spectroscopy. In this dense dispersion of charge-stabilized and magnetic nanoparticles, the interaction potential can be tuned, from quasi-isotropic to anisotropic by applying an external magnetic field. Structural and dynamical anisotropies are reported on interparticle lengthscales associated with highly anisotropic cooperativity, almost two orders of magnitude larger in the field direction than in the perpendicular direction and in zero field.
\end{abstract}

\pacs{Valid PACS appear here}
\maketitle
\par
In many classes of disordered systems, the transition from the fluid state to the solid state remains puzzling : when decreasing the temperature, or increasing the density, the dynamics slows down by several orders of magnitude, whereas static structural properties are weakly affected.  This abrupt slowing-down, referred to as the \emph{glass transition}, is observed in numerous disordered systems \cite{GTReview13,BiroliRevModPhys11}, from frustrated magnets, to molecular glasses, colloidal glasses and granular systems. It is widely accepted that the divergence of the relaxation time is accompanied by a growing length scale resulting in dynamical heterogeneities \cite{BiroliJChem13,ShallScience07,BerthierJCP07,CipPRL09b}. The structural relaxations occur through cooperative rearrangements of clusters of mobile particles. The size of these dynamically correlated clusters can be evaluated, through the computation of dynamical susceptibilities \cite{BerthierJCP07,AbetePRE08}, $\chi_4 \propto N (\left<F^{2}(t)\right>- \left<F(t)\right>^{2})$, with N the number of dynamically correlated particles and $F$, a time-dependent correlator, for example the intermediate scattering function \cite{AbetePRE08}. For hard-sphere systems, the Mode Coupling Theory (MCT) predicts scaling laws \cite{BerthierJCP07,CipPRL09b}, both for the  divergence of the relaxation time and the dynamical susceptibility.
\par
Several recent experimental works have therefore been dedicated to the extraction of the size of dynamically heterogeneous regions:  in molecular and magnetic systems, the measurement of the dynamical susceptibility is delicate and requires an indirect evaluation through non-linear response measurements \cite{BouchaudPRL12,ThibiergePRL10,ElMasriScience06,BertPRL04}. An increase of $\chi_4$ is observed as temperature is lowered.  In colloidal systems, experiments using confocal microscopy \cite{BonnPRL13,ShallScience07} or scattering techniques \cite{CipPRE07,DuriPRL09} provide an easier access to space and time resolved dynamics, and thus to $\chi_4$. In \cite{CipPRL09b} the growth of the dynamical susceptibility with the packing fraction $\Phi$ in colloidal hard sphere glasses is observed.  $\chi_4$ diverges at $\Phi \approx 64 \%$, but MCT predictions do not fit the data over the whole $\Phi$ range. However, fine variations and determinations of $\Phi$ are difficult to perform experimentally in concentrated samples and require many independent experiments. In addition, the relevance of such an averaged macroscopic quantity to characterize the nature of local structural relaxations  is questionable \cite{HawPhilTrans09}. An external parameter allowing to perturbate the local structure at constant $\Phi$ while 
measuring its dynamical consequences would be of great interest experimentally.\\
Precisely to unravel the relationship between the local structure, local dynamics and mechanical properties, micro-rheological experiments and simulations have recently been performed \cite{CandelierPRL09,BocquetPRL11,BinderPRL12}. 
In \cite{BinderPRL12} the authors consider a single particle driven by a constant force in a glass-forming system. They show that beyond the linear response regime, an anomalous -- superdiffusive -- dynamics takes place in the parallel direction to the force, while it remains diffusive in the perpendicular direction. Interestingly, this superdiffusive anisotropic dynamics is associated with wide tails in the distribution of displacements. These features are in agreement with biased trap models \cite{BouchaudRepPhys90} that consider the dynamics of a particle in an anisotropic free energy landscape. Experimental measurements of the dynamical susceptibility would help to understand this anomalous dynamics in biased energy landscapes or in driven systems.

\par In this Letter, we report experiments on a colloidal glass based on magnetic nanoparticles and we study the effect of an external magnetic field on both structural and dynamical properties. We show that if structural anisotropy is weak under field, the dynamics is accelerated and strongly anisotropic: in the direction parallel to the field, we observe giant dynamical susceptibilities associated with shorter relaxation times. We discuss these effects considering the fluctuations induced by local magnetization inhomogeneities along the field direction as an external source of energy.  

\par The investigated system is  a repulsive colloidal glass consisting of charged nanoparticles (NP) of maghemite $\gamma-Fe_2O_3$ (diameter $d\approx$10 nm) dispersed in water. The interparticle repulsion due to the surface charge of these particles dominates and can be tuned with the ionic strength. The glassy sample is prepared from a fluid sample using an osmotic stress technique \cite{MeriguetJPCB06}. This allows to set the ionic strength and to increase $\Phi$ slowly enough to control the state of interaction of the system \cite{WandersJPCM08}. Each NP is also a nanomagnet with a magnetic moment $\mu\sim 10^4 \mu_B$. In absence of magnetic field \textbf{H}, the magnetic moments of the NPs are randomly oriented. The magnetic dipolar interaction contribution, attractive on average, remains small ($U_{dd}\sim 0.7$ $k_{B}T$) with respect to the electrostatic repulsion ($U_{elec}\sim$ 10 $k_{B}T$). The interaction potential is dominated by electrostatic repulsion \cite{MeriguetJPCM05} and in good approximation isotropic. We report here on systems in similar conditions to those of \cite{RobertEPL06} which display slow dynamics and aging above $\Phi=25\%$.


\par The translational dynamics is probed using X-ray Photon Correlation Spectroscopy (XPCS) at the ID10C beamline of the ESRF (Grenoble, France). The energy of the partially coherent incident beam (40 $\times$ 40 $\mu$m$^2$)  is set to 7.03 keV to minimize sample X-ray absorption. The coherently scattered X-rays form speckle patterns on a Charge Coupled Device (CCD) detector (Princeton Instrument), located  3.3 m from the sample. In this configuration, the range of accessible wavevectors ($0.02<Q<0.05 \AA^{-1}$) covers the interparticle lengthscales. Depending on the signal, a speckle pattern is collected every 8 to 15 s to get sufficient statistics. A more precise measure of the static structure (with complementary Small Angle X-ray Scattering, SAXS) is performed at the ID02 beamline of the ESRF at 12 keV. Both experiments are performed with and without an applied field. A thin sample layer ($\sim$ 100 $\mu$m) is deposited on the wall of a quartz capillary, previously filled with dodecane to avoid water evaporation during the experiment. The sample age $t_w$ is initiated as soon as the capillary is placed in measuring conditions and no more subjected to any external mechanical stress. To apply the external permanent magnetic field \textbf{H} (320 kA/m corresponding to 90\% of dipole alignment), a set of magnets is added to the setup, perpendicularly to the beam direction, in the horizontal plane. They are removed for \textbf{H}=0 experiments, to suppress any residual field. Note that the colloidal dispersion remains monophasic in presence of a \textbf{H} field \cite{MeriguetJPCB06}. Last, under-field the scattering pattern is anisotropic. The intensity is thus analyzed over angular sectors of 20\degre both along or perpendicularly to the field direction. 

\begin{figure}[!t]
\includegraphics[width=7cm, trim =1cm 0.2cm 0cm 0cm, clip]{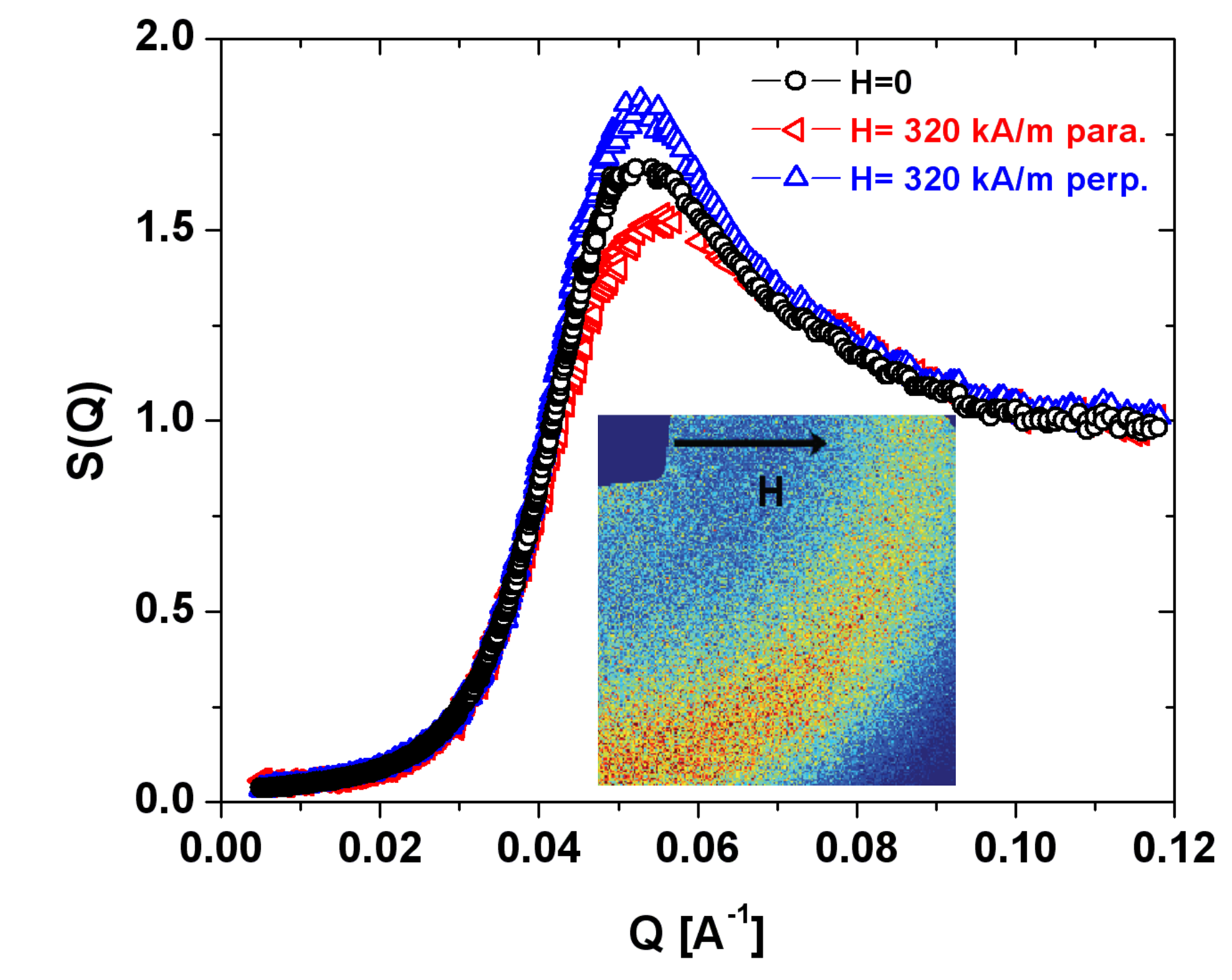}
\caption{\label{Fig1} Static structure factor $S(Q)$ obtained by SAXS, for \textbf{H}=0 (circles) and \textbf{H}=320 kA/m (triangles). Inset: speckle pattern of a quadrant of the scattered intensity for \textbf{H}=320 kA/m. The direction of \textbf{H} is indicated by the arrow.}
\end{figure}

We investigated the dynamics of a sample with a volume fraction $\Phi=30\%$. It is a freshly prepared glass-former \cite{WandersRot09} chosen for its slow dynamics, lying in the accessible time window of XPCS. We do not observe any noticeable evolution of its static structure on this time window. The structure factor S(Q) without field (Fig. \ref{Fig1}) determined by SAXS is characteristic of repulsive interactions, with a low compressibility of 0.04 and a mean distance of $\sim$ 12 nm between the NPs. Each NP can be seen as an effective sphere of diameter $d+2\kappa^{-1}\approx 12.5$ nm, where $d$ is the solid diameter of the NP and $\kappa^{-1} \approx 1.2$ nm, the effective screening length of the potential. This latter is evaluated by adjusting the $\Phi$-dependence of the compressibility, using the Carnahan-Starling formalism \cite{WandersSoftMat13}. For the studied sample at $\Phi=30\%$, the mean interparticle distance is comparable to the size of the effective spheres. They are sligthly interpenetrated, giving rise to the glassy behavior of the sample.

\par When applying a magnetic field (\textbf{H}=320 kA/m), the magnetic dipoles align along \textbf{H}. Since interparticle repulsion is dominant, the solid bodies of the NPs remain well apart from each other and there is no signs of self-assembled anisotropic structures. However the interaction potential becomes slightly anisotropic, inducing a slightly anisotropic structure illustrated here by the anisotropy of the scattered intensity and S(Q) (Fig. \ref{Fig1}). The largest anisotropy, observed on the scale of the interparticle distance, i.e. on the peak of S(Q), is quantified by its maximum $S_{max}$ and similar to the behavior in ferrofluids in the fluid state. $S_{max}$ increases in the perpendicular direction, due to the increase of interparticle repulsion under magnetic dipolar interaction. $S_{max}$ decreases in the parallel direction due to the effect of dipolar interaction combined with NPs fluctuations along the field direction around their most probable position \cite{GazeauPRE02, MeriguetJPCB06}. Only a small shift of its position $Q_{max}$ occurs while, on larger scales, the compressibility remains almost isotropic. 

We already have investigated the anisotropy of the local structure over the whole $\Phi$ range in a recent paper \cite{WandersSoftMat13}, showing that the NPs are trapped in an anisotropic cage formed by their first neighbors.  The structural anisotropy is interpreted by a model of magnetostriction, lying on the under-field deformation of anisotropic cages formed by neighboring NPs.  This model relies on the elastic deformation of the cage at constant volume under the applied field and catches the $Q_{max}$ and $S_{max}$ anisotropies. It provides a measure of the elastic modulus of the
magnetic fluid. In this model the anisotropy of the cage writes \cite{WandersSoftMat13}:
\begin{equation}
\alpha = \frac{d_{\perp}-d_{//}}{d_0}=\frac{9}{4\pi}\frac{\mu_0 M_{MF}^2(H) \sigma_H^2}{k_BT}
\end{equation}
\noindent where $d_{\perp}$ (resp. $d_{//}$) is the dimension of the cage in the perpendicular (resp. parallel) direction of the field, $d_0$ its dimension in absence of field, $M_{MF}(H)$ the magnetization of the magnetic fluid and $\sigma_H^2$ the mean squared displacement of the NP (assumed isotropic in the model). The anisotropy of the cage is found to be moderate $\alpha \approx 8 \%$ \cite{WandersSoftMat13} but means that the interpenetration of the effective spheres under an \textbf{H} field is slightly anisotropic. This anisotropic interpenetration could carry dynamical consequences. Interestingly, the widening  of S(Q)  in the parallel direction suggest $\sigma_H^2$ to be anisotropic with $\sigma_{\perp} < \sigma_\parallel$. Only XPCS dynamical study can shed some light on this question. 

\par From the XPCS measurements we compute the degree of temporal correlation between two speckle patterns, separated by a lag-time $\tau$, at a time $t$ \cite{DuriEPL06,CipPRE07}:
\begin{equation}
c_{I}(Q,\tau,t) = \frac{\left\langle I_{p}(Q,t)I_{p}(Q,t+\tau)\right\rangle_{Q}}{\left\langle I_{p}(Q,t)\right\rangle_{Q} \left\langle I_{p}(Q,t+\tau)\right\rangle_{Q}} - 1
\label{EqCi}
\end{equation}
\noindent where $I_{p}(Q,t)$ is the scattered intensity at pixel $p$ and time $t$, $\left\langle\dots\right\rangle_Q$ denotes an ensemble average over pixels in a ring of iso-wavevectors $Q$ (for isotropic patterns). Under field, the scattered intensity is anisotropic, thus the average in Eq.(\ref{EqCi}) is restricted to the same angular sectors as for $S(q)$ along and perpendicular to the field direction. We also observe that for zero field (i.e. isotropic case) this average on a reduced azimutal area does not influence the obtained results. The time average of $c_I$ provides the usual intensity autocorrelation function:
\begin{equation}
g^{(2)}(Q,\tau,t_{w}) = \left\langle c_{I} (Q,t,\tau)\right\rangle_{t_w}+ 1
\label{g2Eq}
\end{equation}
\noindent where the brackets $ \left\langle\dots\right\rangle_{t_w}$ denote a time average (over 200 frames) centered on age $t_w$ \footnote{For the present sample, the age dependence of the dynamics is complex as two aging regimes are observed \cite{RobertEPL06} for H=0 and an anisotropic aging is observed under field (data not shown). The $c_I$'s are analyzed here in the slow aging regime of \cite{RobertEPL06}.}. In addition, we compute the temporal variance of $c_I$, normalized as described in \cite{WandersJPCM08,CipNaturePhys08}. It is related to the dynamical susceptibility $\chi_{4}$, the volume integral of a 4-point correlator \cite{CipNaturePhys08,BerthierJCP07,BiroliJChem13,ThibiergePRL10}. This dynamical susceptibility, hereafter referred to as $\chi(\tau)$ is calculated at all accessible $Q$'s and provides information on the heterogeneous nature of the dynamics.\par  
%
\begin{figure}[!t]
\includegraphics[width=8.5cm,trim = 2cm 1cm 3cm 2cm, clip]{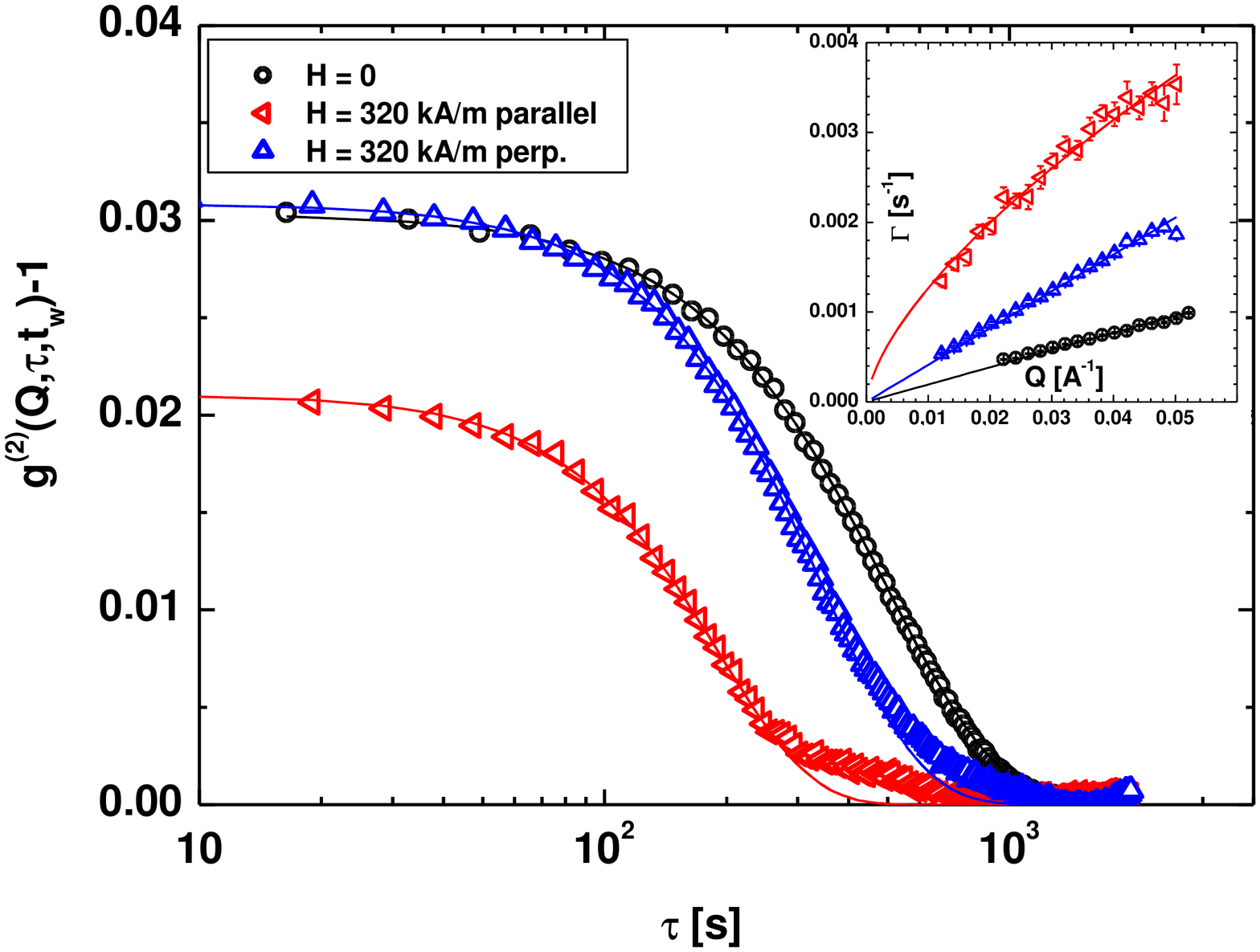}
\caption{\label{Fig2} Intensity autocorrelation function $g^{(2)}(Q,\tau,t_{W})$ as a function of $\tau$ for $Q$=0.048 $\AA^{-1}$ and $t_w$=5622 s. Solid lines are compressed exponential fits, $g^{(2)}-1 \propto \mathrm{exp}\left[-2\left(\Gamma t\right)^{\beta}\right]$, with $\beta>1$. Inset : $Q$-dependence of $\Gamma$, at $t_w$=5622 s. Full lines are linear fits at low $Q$'s, for H=0 and in the parallel direction. For the perpendicular direction, the full line is a power law fit $\Gamma=0.025~Q^{0.66}$.}
\end{figure}
%
The intensity autocorrelation functions at $Q$=0.048 $\AA^{-1}$ and $t_w$=5622 $\pm$375 s, are plotted in Fig.~\ref{Fig2} with and without applied field. The observed decay of $g^{(2)}$ is faster when a field is applied, this effect being stronger along the field direction and associated to a smaller non-ergodicity factor (defined here as $g^{(2)}$ at short times $\tau$). These features are observed on all probed $Q$'s and $t_w$'s.
The $g^{(2)}$'s can be modelled by compressed exponentials \cite{RobertEPL06} and the inverse characteristic time $\Gamma=\tau_c^{-1}$ increases linearly with $Q$ for $H$=0, as observed on many glassy systems \cite{CipNaturePhys08}. In presence of a \textbf{H} field this ballistic-like behavior is only observed in the perpendicular direction (see inset of Fig.~\ref{Fig2}). In the parallel direction, we observe a non linear $\Gamma(Q)$ relationship, better noticeable at high $Q$'s. This behavior is reminiscent of de Gennes' narrowing, occuring at lengthscales close the structure factor peak \cite{CaronnaPRL08}. This non linearity is also comparable to the superballistic behavior evidenced in \cite{DuplatPRE13} with $\Gamma\sim Q^{2/3}$ (inset of Fig.~2). However, given the limited $Q$ range of our XPCS experiment we cannot properly conclude on the origin of this non linearity. Interestingly however, such an anomalous dynamics in the direction of the field is reported in theoretical works that consider the anisotropic dynamics of a single particle either driven by a constant force in a glass-forming system \cite{BinderPRL12} or evolving in a biased potential energy landscape \cite{BouchaudRepPhys90}. For moderate values of the driving force, the dynamics in the parallel direction to the force is superdiffusive -- whereas it remains diffusive in the perpendicular direction -- and this behavior is associated to a wide-tail distribution of displacements. In our system, the application of a \textbf{H} field tends to reduce the energy barriers in the direction of the field, due to dipolar interactions. This biased and anisotropic energy landscape could thus be at the origin of such an anomalous $\tau(Q)$ relationship in the parallel direction. If so, one would expect large temporal fluctuations of the dynamics in the parallel direction and thus a possible signature of it in the dynamical susceptibilities.
\par
\begin{figure}[!t]
\includegraphics[width=9cm]{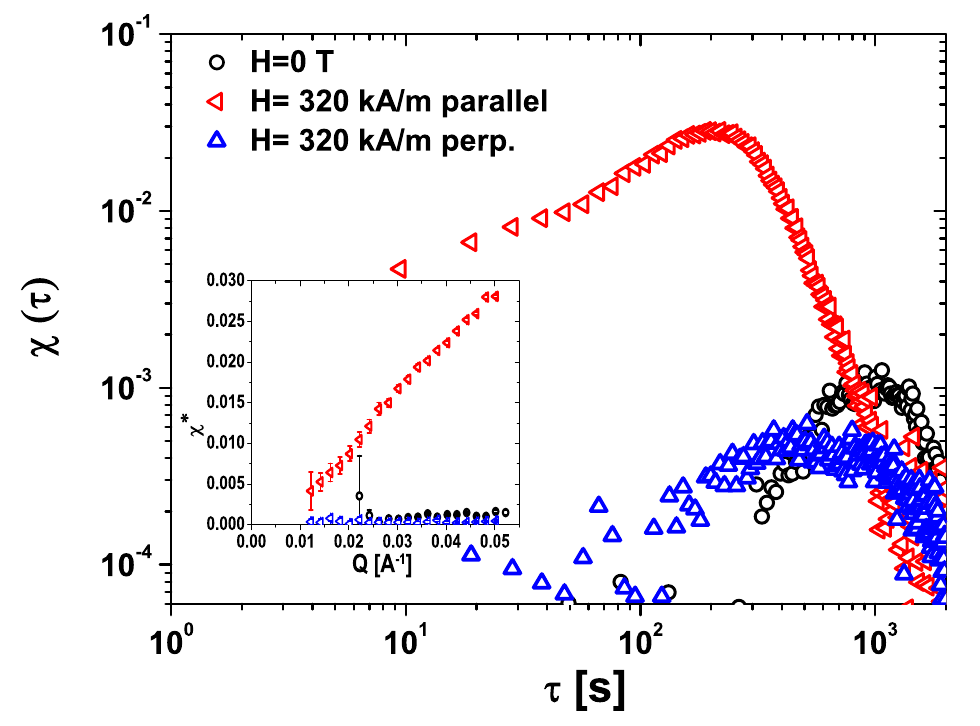}
\caption{\label{Fig3} Dynamical susceptibilities $\chi(\tau)$ at $Q=0.048\AA^{-1}$, at 0 and 320 kA/m, parallel or perpendicular to the field direction. Inset: maximum of $\chi(\tau)$ as a function of $Q$ }
\end{figure}

 The dynamical susceptibility $\chi(\tau)$ without and with magnetic field is plotted in Fig.~\ref{Fig3} for $Q$=0.048 $\AA^{-1}$. When a dynamical cluster of NPs relaxes in the sample, it induces a drop of the correlator $c_I$. $\chi(\tau)$ is then maximum on the timescale of these relaxation processes and its maximum $\chi^*$ is a measure of the degree of cooperativity of the dynamics \cite{WandersJPCM08,CipNaturePhys08}. Therefore the peak position corresponds to the characteristic time $\tau_c$ of $g^{(2)}$ and reproduces the anisotropy seen in Figure \ref{Fig2}. The $\chi^*$ values are also highly anisotropic being almost two orders of magnitude larger in the parallel direction than in perpendicular one. Assuming that $\chi^*$ is proportionnal to the cube the size of the dynamic clusters \cite{CipNaturePhys08}, it gives dynamically correlated clusters elongated along the field with an anisotropy ratio of the order of 4 to 5.

\par The $Q$-dependence of $\chi^{*}$ is presented in the inset of figure \ref{Fig3}. When $H=0$, $\chi^{*}$ is nearly constant with $Q$, in agreement with previous results of \cite{WandersJPCM08}. Under field, $\chi^{*}$ in the perpendicular direction is even smaller than for $H=0$, and characterizing its $Q$-dependence is difficult. In the parallel direction $\chi^{*}$ grows linearly with $Q$, providing evidence that the cooperative process takes place on local length-scales. The faster dynamics in the parallel direction appears to be driven by cooperative events taking place at short time/lengthscales.


\par The application of a magnetic field on this repulsive and magnetic colloidal glass has both structural and dynamical consequences. We can first interpret the field effect as a simple geometrical perturbation of the zero field situation. Since the magnetic dipolar interactions reduce the repulsion between particles in the direction of the field, the system can still be considered as effective spheres --  soft and interpenetrated -- but trapped in \emph{anisotropic cages} and coupled to both an external magnetic field and to a thermal bath. There is a growing interest to consider the dynamics of anisotropic systems, such as hard ellipsoids \cite{ZhengPRL12} or dumbbells \cite{KobPRL09}. For instance, Zheng \emph{et al.} measure the dynamics of concentrated hard ellipsoids dispersions \cite{ZhengPRL12}. They show that at a given $\Phi$, the cooperative nature of the translational dynamics is anisotropic. However, the situation we consider here is different in essence, since the anisotropy of the interaction is due to a global coupling to the \textbf{H} field rather than set by a local anisotropic range of interaction as for hard ellipsoids.  The coupling to the field modulates the local fluctuations of position of the NPs. These fluctuations are anisotropic, being larger along the field direction and are hindered by the cages inside which the NPs are confined. They are then the ``motors" of coherent fluctuations of adjacent cages in the direction of the field, increasing the cooperativity and decreasing the non-ergodicity factor on the scale of a few interparticle distances.


\par Consequently, we can deduce that in presence of the field, the complex energetic landscape in which the system evolves is anisotropic. Due to the reduced repulsion along the field direction, the time scale of barrier-crossing is lower and the relaxation is faster.  In qualitative agreement with \cite{BinderPRL12,BouchaudRepPhys90} we report an anomalous dynamics $\tau(Q)$ in the parallel direction to the field, together with highly cooperative dynamics, due to the field-driven fluctuations of position of the NPs. This enhanced cooperativity in the direction of reduced repulsion can be compared to  simulations of hard spheres where the interaction potential is tuned \cite{ConiglioJPCM07,ConiglioJPCM09}. By computing numericallly $\chi_4$ in (isotropic) colloidal hard sphere glasses and gels, Coniglio \emph{et al.} show that in glasses  $\chi_4$ is peaked in time, traducing the transcient nature of cooperative process, whereas in gels $\chi_4$ reaches a plateau, traducing the persistent presence of relaxing clusters containing a large number of particles. In our case, in the parallel direction to \textbf{H}, the system is not made of permanent structures at contact, the rearrangements of which lead to a giant cooperativity on long timescales. On the contrary, $\chi(\tau)$ we report on seems to plateau at short times, the dynamical fluctuations being driven by short timescales field induced fluctuations.\\
In the perpendicular direction, the dynamics is slower \emph{and} less cooperative. This is in contradiction with the picture of growing cooperative domains, which usually explains the slowing down of the dynamics in glasses. However, the reduced particle displacements in very dense suspensions tends to decrease $\chi^{*}$ despite of the growing size of the cooperative clusters \cite{CipNaturePhys08}. It leads to a non-monotonic $\Phi$-dependence of $\chi^{*}$, with a maximum at a volume fraction $\Phi_c$ and a decay for higher $\Phi$'s. A similar scenario could take place here, supported by the observed reduced mean squared displacements in the perpendicular direction $\sigma_{\perp} < \sigma_0 < \sigma_\parallel$ \cite{WandersSoftMat13}, along with the widening of $S(Q)$ around $Q_{max}$ in the parallel direction.
\par In conclusion, using XPCS experiments, we have probed the translational dynamics of a repulsive colloidal glass made of magnetic nanoparticles. Under the application of a \textbf{H} field the dynamics of the system is faster than in zero field. The dynamics is anisotropic, being faster along the field than in the perpendicular direction. In the parallel direction an anomalous dynamics $\tau(Q)$ is observed, associated to a huge increase of the dynamical susceptibility, in qualitative agreement with biased trap models \cite{BouchaudRepPhys90} and recent numerical simulations \cite{BinderPRL12}. All these features are qualitatively recovered at any ages of the sample. The field intensity could thus be used as a way to tune the dynamical cooperativity, and thus to study its eventual divergence in the parallel direction.
This system, which catches the general features of the glass transition, have through the magnetic field a powerful control parameter. Conjugating experiments on glassy samples at various $\Phi$ probed at various $H$ is a promising route to understand how the dynamical heterogeneities vary at the approach of the glass transition.\\

\begin{acknowledgments}
We acknowledge P. Panine for his support on ID02 beamline in ESRF for SAXS measurements and F\'ed\'eration F21 UPMC-Paris 6 for financial support.
\end{acknowledgments}
\bibliographystyle{unsrt}
\bibliography{apssamp}
\end{document}